\documentclass[12pt]{iopart}

\usepackage{iopams}  
\usepackage{setstack}  

\usepackage{iopams}  
\usepackage{setstack}  

\usepackage{graphicx}% Include figure files
\usepackage{dcolumn}% Align table columns on decimal point

\newcommand{\myskip}[1]{}

\newcommand{\vf}{{\bf f}}
\newcommand{\vq}{{\bf q}}
\newcommand{\vu}{{\bf u}}
\newcommand{\vv}{{\bf v}}

\newcommand{\ccdot}{\hspace{-1mm}\cdot\hspace{-1mm}}

\newcommand{\vB}{{\bf B}}
\newcommand{\vE}{{\bf E}}

\newcommand{\vA}{{\bf A}}
\newcommand{\vC}{{\bf C}}

\newcommand{\vL}{{\bf L}}

\newcommand{\cE}{{\cal E}}

\newcommand{\vk}{{\bf k}}
\newcommand{\vn}{{\bf n}}
\newcommand{\vp}{{\bf p}}
\newcommand{\vr}{{\bf r}}
\newcommand{\vs}{{\bf s}}

\newcommand{\vnul}{{\bf 0}}

\newcommand{\B}{{\bf B}}

\newcommand{\BEQ}{\begin{eqnarray}}
\newcommand{\EEQ}{\end{eqnarray}}
\newcommand{\BEA}{\begin{eqnarray}}
\newcommand{\EEA}{\end{eqnarray}}

\newcommand{\nn}{\nonumber }
\renewcommand{\d}{{{\rm d}}}

\newcommand{\eps}{\varepsilon}
\newcommand{\om}{\omega}

\newcommand{\half}{\frac{1}{2}}

\begin{document}

\title[The H ground state in SED] {%SED under the test: 
Simulation of the hydrogen ground state in Stochastic Electrodynamics}
 
\author{Theo M. Nieuwenhuizen$^{1,2}$ and Matthew T. P.  Liska$^{1}$}

\address{
$^{1}$ Institute for Theoretical Physics, University of Amsterdam, % Science Park 904, 
P.O. Box 94485, 1098 XH Amsterdam, the Netherlands \\
$^2$ International Institute of Physics, UFRG,
%  Federal University of Rio Grande do Norte,
% Av. Odilon Gomes de Lima, 1722, % - Capim Macio - 
 Av. O. Gomes de Lima, 1722, % - Capim Macio - 
59078-400 
Natal-RN, Brazil
\\
}

\ead{t.m.nieuwenhuizen@uva.nl}

\begin{abstract}
Stochastic electrodynamics is a classical theory which assumes that the physical vacuum 
consists of classical stochastic fields with average energy $\half \hbar\omega$ in each mode, i.e., the zero-point Planck spectrum.
While this classical theory explains many quantum phenomena related to harmonic oscillator problems,
hard results on nonlinear systems are still lacking. In this work the hydrogen ground state is studied by numerically solving the
Abraham -- Lorentz equation in the dipole approximation. First the stochastic Gaussian field is represented by a
sum over Gaussian frequency components, next the dynamics is solved numerically using OpenCL.
The approach improves on work by Cole and Zou 2003 by treating the full $3d$ problem and reaching longer simulation times.
The results are compared with a conjecture for the ground state phase space density.
Though short time results suggest a trend towards confirmation,  in all attempted modelings the atom ionises at longer times.
\end{abstract}

% \pacs{xxx}
 \pacs {11.10, 05.20, 05.30, 03.65}
 
%Uncomment for PACS numbers title message
%\pacs{00.00, 20.00, 42.10}
% \pacs {11.10,05.20,05.30,03.65}
 % Keywords required only for MST, PB, PMB, PM, JOA, JOB? 
%\vspace{2pc}
%\noindent{\it Keywords}: Article preparation, IOP journals
% Uncomment for Submitted to journal title message
%\submitto{\JPA}
% Comment out if separate title page not required

\noindent
Keywords: {Stochastic electrodynamics, hydrogen ground state, numerical simulation}

\maketitle

arXiv:1407.7030

%\pacs{ 04.70.Bw,04.20.Cv, 04.20.Jb}
%\pacs{04.70.Bw}{Classical black holes}
%\pacs{04.20.Cv}{Fundamental problems and general formalism}
%\pacs{04.20.Jb}{Exact solutions}
%\keywords{Black hole, interior solution}

\section{Introduction}

The theory called Stochastic Electrodynamics (SED) starts with a classical picture of what is normally called the quantum vacuum:
the vacuum is assumed to consist of fluctuating classical electrodynamic fields with energy per eigenmode equal to $\half\hbar\om$, which adds up to the 
zero-point Planck spectrum $\rho(\omega)=\hbar \om^3/2\pi^2c^3$. Particles are considered as classical too, hence in the hydrogen problem the electron is 
a point particle that essentially goes around the nucleus in Kepler orbits.  Like any accelerated classical charge it radiates, hence the energy loss causes 
it to fall onto the nucleus, the old problem of the classical atomic model.
The assertion of SED is that this energy loss is statistically compensated by energy gained from the fluctuating vacuum fields, so that the
stability of the hydrogen atom, and more generally of matter, is achieved. 

SED has enjoyed popularity in the seventies and eighties of last century, when many 
linear problems (harmonic oscillator problems) could be reproduced from this classical approach \cite{delaPenaCettobook,Puthoff1987}.
The field lost attention when it became clear that nonlinear problems, such as the hydrogen stability, could not be explained. 
For instance, from a Fokker-Planck analysis it was concluded that the electron would evaporate, thus self-ionising the H atom \cite{delaPenaCettobook}.
Outside the field, the theory is considered as suspicious due to the supposed road block for hidden variables theories by Bell inequalities.
The latter will not be our concern, since one of us has joined a growing group of researchers who are convinced that Bell 
had to make a hidden, unnatural assumption to derive his inequalities, a problem related to the context (setup of detectors).\footnote{
Although Bell assumed that different contexts can be combined, this is not true in general, 
hence it leads to the {\it contextuality loophole},  which cannot be closed for it being a theoretical problem \cite{TheoBellFoP}.
Violation of the Bell inequality demonstrates that the combination of contexts is not allowed, without any
further implication on presence or absence of local realism.}
So the issue of Bell inequality violations should not be held against SED.

Not withstanding the above and other apparent setbacks, several people have continued to develop SED.
de la Pe\~na and Cetto wrote a book \cite{delaPenaCettobook} on it in 1996 and a second one \cite{delaPenaCettoValdesbook},
 with Valdes-Hernandes, was recently published.
They have formulated both a Heisenberg and a Schr\"odinger approach arriving at the familiar equations of QM.
They also consider the problem of entanglement, it being carried by the stochastic field \cite{dlPena2012,Cetto2012};
this seems akin to the creation of two polarons where lattice distortions (phonons) move with the two electrons.
Fran\c{c}a et al. also derive the Schr\"odinger equation and stress the role of the ZPF in the uncertainly relations and 
the photo-electric effect \cite{Franca2012}. One of us considers in SED and SED-like theories 
a ``pull back'' mechanism to turn classical scattering into quantum scattering \cite{TheoPullBack};
the phase space densities of various states of the relativistic hydrogen atom \cite{NRelHatom06};
and an arrow of time: 
The involved energy current (fluctuation energy in, radiation energy out) would define the {\it subquantum arrow of time},
intimately connected to the stability of matter. It is more fundamental than the entropic and cosmological arrows of time \cite{NVienna2014}.

In view of these aspects, we see it as crucial to test SED on a nonlinear problem. 
The most obvious case is the hydrogen ground state.
While unstable in the Fokker-Planck approximation \cite{ClaverieSoto1982,delaPenaCettobook}, the treatment of 
de la Pe\~na and Cetto points to resonances. In the diagrammatic approach of the Liouville equation by \cite{PenaCetto1977}, one of us 
noticed that higher order corrections in the fine structure constant achieve power laws in time, so that 
in the long time limit neither stability nor instability is obvious \cite{NhUnpub2013}.

Lacking analytical derivations it would be desirable to find results from numerical analysis, the first target being the hydrogen ground state. 
This challenge is taken up in 2003 by Cole and Zou \cite{ColeZou2003}.
They discretize the stochastic field and follow the perturbed Kepler orbits up to 7.252 ps, that is, up to 300,000 Bohr periods.
Taking the average over 11 simulated trajectories, they establish an encouraging fit of the radial ground state density. 

With computing power having strongly increased during last decade, it seems appropriate to redo the simulations.
We take up this challenge and compare the results with the conjecture for the phase space density of the ground state \cite{NRelHatom06}.
This theory will be recalled in section 2. 
In section 3 we recall the conjecture for the phase space density and express this as a conjecture for the distribution of conserved
quantities energy and angular momentum. 
The simulation results will be reported in section 4. 
We close with a discussion. The appendix gives some details about our code in OpenCL.

\section{The hydrogen problem in stochastic electrodynamics}

The Newton equation for the electron with damping and noise, also called the Abraham-Lorentz equation or Brafford-Marshall equation, reads

\BEQ
m\ddot\vr=-\frac{Ze^2\vr}{4\pi\epsilon_0r^3}+\frac{e^2}{6\pi\epsilon_0c^3}\,\dddot \vr -e[\vE(\vr,t)+\dot\vr\times\vB(\vr,t)]
\EEQ
where $\vr=\vr(t)$ is the orbit and we stress the explicit time-dependence of $\vE$ and $\vB$.
The first term on the right hand side is the Coulomb force on a charge $-e$ by a central charge $Ze$, 
the second the damping term, which arises together with a renormalisation of the mass, so that $m$ 
is the physical mass of the electron. 

The conserved quantities of the unperturbed problem are the energy, angular momentum 
and the Lagrange-Runge-Lenz eccentricity vector,

% \end{document}

\BEQ 
\hspace{-2cm}
\cE=\frac{\vp^2}{2m}-\frac{Ze^2}{4\pi\epsilon_0r},\qquad \vL=\vr\times\vp,\quad
% {\eps}=\frac{4\pi\epsilon_0}{Ze^2m}\vp\times\vL-\hat\vr.
\bvarepsilon=
\frac{1}{m}\vp\times\vL-
\frac{Ze^2}{4\pi\epsilon_0}\hat\vr.
\EEQ

The vector potential and the electric and magnetic fields are sums of plane waves with random coefficients

\BEA
\vA&=&
%\frac{1}{\sqrt{L_1L_2L_3}}
\sum_{\vk,\lambda}
%e^{-\hbar\om_\vn/mc^2}
%\sqrt{2\pi\hbar\om_\vn}
\sqrt{\frac{{{\cal E}_\vn}}{\epsilon_0L_xL_yL_z}}\,
\frac{\hat\eps_{\vn\lambda}}{\omega_\vn}[A_{\vn\lambda}\sin(\vk \cdot\vr-\om_\vn t)
+B_{\vn\lambda}\cos(\vk \cdot\vr-\om_\vn t)] \nn\\
 \vE&=&
 %\frac{1}{\sqrt{L_1L_2L_3}}
\sum_{\vk,\lambda}
%e^{-\hbar\om_\vn/mc^2}
%\sqrt{2\pi\hbar\om_\vn}
\sqrt{\frac{{{\cal E}_\vn}}{\epsilon_0L_xL_yL_z}} \, \label{Esum}
\hat\eps_{\vn\lambda}[A_{\vn\lambda}\cos(\vk \cdot\vr-\om_\vn t)
-B_{\vn\lambda}\sin(\vk \cdot\vr-\om_\vn t)] 
%\qquad \epsilon_0\to\frac{1}{4\pi}
\\
 \B&=&
%\frac{1}{\sqrt{L_1L_2L_3}}
\sum_{\vk,\lambda}
%e^{-\hbar\om_\vn/mc^2}
%\sqrt{2\pi\hbar\om_\vn}
\sqrt{\frac{{\mu_0{\cal E}_\vn}}{L_xL_yL_z}}\,\hat \vk\times
\hat\eps_{\vn\lambda}[A_{\vn\lambda}\cos(\vk\cdot\vr-\om_\vn t)
-B_{\vn\lambda}\sin(\vk\cdot\vr-\om_\vn t)] \nn \EEA
Since we adopt periodic boundary conditions, the wave vector components $k_a=2\pi n_a/L_a$ involve integer $n_a=-\infty, \cdots,-1,0,1,\cdots,\infty$, ($a=x,y,z$).
The $\hat\eps_{\vn\lambda}$ with ($\lambda=1,2$) are polarisation vectors.
The $A_{\vn\lambda}$ and $B_{\vn\lambda}$ are independent random Gaussian variables with average zero and unit variance.
For each term the energy  $\int_V\d^3r(\frac{\epsilon_0}{2}\vE^2+\frac{1}{2\mu_0}\vB^2)$ is in integral equal to $\cE_\vn$, for which we
 choose the zero point energy combined with an exponential cutoff at the electron zero point energy, $\cE_\vn=\half\hbar\om_\vn\,\exp(-\hbar\om_\vn/mc^2)$.
The correlation function of the stochastic electric field is translation invariant in space and time. We shall need

\BEQ
C_{ij}^{EE}({\bf 0},t)=\langle E_i(\vr,t)E_j(\vr,0)\rangle = 
\delta_{ij}\frac{\hbar}{\pi^2\epsilon_0c^3}\,\Re\frac{1}{(t+i\hbar/mc^2)^4}             
\EEQ
For our application to the H atom, we go to Bohr units, 

\BEQ 
%a_0=\frac{\hbar}{Z\alpha mc} %=\frac{a_B}{Z} ,\qquad \tau_0=\frac{1}{\om_0}=\frac{\hbar}{Z^2\alpha^2 mc^2}
%=\frac{\tau_B}{Z^2},\qquad 
a_0=\frac{\hbar}{Z\alpha mc} %=\frac{a_B}{Z}
,\qquad \tau_0=\frac{1}{\om_0}=\frac{\hbar}{Z^2\alpha^2 mc^2}
\EEQ
$\tau_0$ is the characteristic Bohr time and  the Bohr period is $P_0=2\pi\tau_0$.
In these units the equation of motion becomes

\BEQ \label{BMar}
\ddot\vr=-\frac{\vr}{r^3}+\beta^2 \dddot{\vr} -\beta[\vE(Z\alpha\vr,t) +Z\alpha\dot\vr\times\vB(Z\alpha\vr,t)],\qquad 
\EEQ
% \BEQ \ddot\vr(t)=-\frac{\vr(t)}{r^3(t)}+\beta^2 \dddot\vr(t) -\beta\{\vE[\vr(t),t] +Z\alpha\dot\vr(t)\times\vB[\vr(t),t]\},\qquad \EEQ
Both the fluctuations and the damping involve the small parameter 
\footnote{In order to have $\beta$ also as prefactor of $\vE$, we absorb a factor $\sqrt{3/2}$ in  $\vA$, $\vB$ and $\vE$.\label{fntd}}
\BEQ \label{beta=}
\beta=\sqrt{\frac{2}{3}}Z\alpha^{3/2}=\frac{Z}{1964.71}
, \qquad  \alpha=\frac{e^2}{4\pi\epsilon_0\hbar c}\approx\frac{1}{137}
\EEQ
It is seen that the effect of $Z>1$ is to make the fluctuations and damping stronger, suggesting a speed up in the simulations. 

The phase $\vk\cdot\vr-\om t$ of the plane waves of the EM fields reads in Bohr units $Z\alpha \vk\cdot\vr-kt$, so 
to leading order we may neglect the spatial dependence of the electric field (dipole approximation),
while we can also omit the magnetic field. 
Now $\vE(t)\equiv\vE(\vnul,t)$ is given as in (\ref{Esum}) at $\vr={\bf 0}$, with the argument of the square root replaced by the dimensionless expression
$3\pi\,\om \tau_0\,(c\tau_0/L_a)^3$.
% is  $\sqrt{3\pi\tilde \om_\vn/\tilde L^3}$ where $\tilde\om=\om\tau_0$ and $\tilde L=L/c\tau_0$.
The autocorrelation function thus reads

\BEQ
\label{CEEexact}
C_{ab}^{EE}(t)= %-\frac{1}{c^2}\ddot C_{AA}(t)=
\langle E_a(t)E_b(0)\rangle= 
\delta_{ab}\frac{6}{\pi}\,\Re\frac{1}{(t-iZ^2\alpha^2)^4}  ,           
\EEQ
with $t$ expressed  in Bohr times. 
% The factor $6$ stems from the factor $\frac{2}{3}$ in  the definition  (\ref{beta=})  of $\beta$.

After iterating the damping term in Eq. (\ref{BMar}), we arrive at

\BEQ \label{neqad0}
\ddot\vr=-\frac{\vr}{r^3}-\beta^2\frac{\dot \vr-3(\dot\vr\cdot\hat\vr)\hat \vr}{r^3} -\beta\vE(t), %\qquad \beta=\sqrt{\frac{2}{3}}Z\alpha^{3/2}
\EEQ
which we may write as

\BEQ
\label{neqad}
\dot\vp=\vf(\vr) -\beta^2\dot\vf   -\beta\vE(t), \qquad \dot\vr=\vp,\qquad
\vf(\vr)=-\frac{\vr}{r^3}
\EEQ
where $\dot\vf \equiv \nabla \vf(\vr)\ccdot\dot \vr$.
The conserved quantities at $\beta\to0$ now read 

\BEQ 
\cE=\half\vp^2-\frac{1}{r},\qquad \vL=\vr\times\vp,\qquad 
\bvarepsilon =p^2\vr-(\vp\cdot\vr)\vp-\hat\vr
\EEQ
It follows that

\BEQ \label{epsEL}
\eps^2=1+2\cE L^2.
\EEQ
The relation %taking the inner product with $\vr$ yields 
$\bvarepsilon \cdot \vr=L^2-r$ can be expressed as

\BEQ
\label{rphi}
% \bvarepsilon \cdot \vr=L^2-r,\qquad 
r=\frac{L^2}{1+\eps\cos\phi}
=\frac{(1-\eps^2)R}{2(1+\eps\cos\phi)},\qquad R\equiv -\frac{1}{\cE}>0,
\EEQ
where $\phi$ is the angle between $\bvarepsilon $ and $\vr$.
Thus $\eps$ expresses the eccentricity of the orbit.

\subsection{Simplified representation of the stochastic field}

The stochastic electric field  involves the numerically demanding   $3d$ sum over $\vk$ values.
%  and polarisation vectors.
To facilitate the simulations, we replace it by a simpler Gaussian field.
% obtained by discretising (\ref{CEEexact}). 
We adopt a uniform grid in $\om$-space with $\Delta\om_n=1/N$ with $N\gg1$, so that 

\BEQ
\om_n=\frac{n}{N},\qquad \qquad (n=1,2,\cdots),\qquad 
% \Delta\om_n=\frac{1}{N},
\EEQ
% for some large $N$, 
which corresponds to $({n}/N)\, \om_0$ in physical units.
Next we assume for each $n$ and for each direction $a=x,y,z$, two independent Gaussian random variables $A_n^a$ and $B_n^a$, with 
average 0 and  variance 1, and consider the $1d$ sum

%\BEQ\label{EadtB} \vE(t)=\sum_{n=0}^\infty \sqrt{\frac{n^3}{\pi N^4}} e^{-Z^2\alpha^2n/2N} (\vA_n\cos\frac{nt}{N}+\vB_n\sin\frac{nt}{N}).
 % w_n=\sqrt{\frac{n^3}{\pi N^4}\EEQ

\BEQ\label{EadtB}
\vE(t)=\sum_{n=0}^\infty \sqrt{\frac{\Delta\om_n\,\,\om_n^3}{\pi}} e^{-\half Z^2\alpha^2\om_n} 
(-\vA_n\cos\om_nt+\vB_n\sin\om_nt).
 % w_n=\sqrt{\frac{n^3}{\pi N^4}
\EEQ
Its two-point correlation function reads

\BEQ
\hspace{-2cm}
C_{ab}^{EE}(t-s)=\delta_{ab}C_{EE}(t-s),\quad
C_{EE}(t)=\frac{1}{8\pi N^4}\, \Re\, \frac{3+\sinh^2[(Z^2\alpha^2+it)/2N]}{\sinh^4[(Z^2\alpha^2+it)/2N]}.
\EEQ
At fixed $t$ it reproduces in the limit $N\to\infty$ the autocorrelation function (\ref{CEEexact}).
For finite $N$, the discretization will be reliable for times up to $t\simeq N$.

The related $\vA$ field reads

\BEQ\label{AadtB}
\vA(t)=\sum_{n=1}^\infty \sqrt{\frac{\Delta\om_n\,\,\om_n}{\pi}} e^{-\half Z^2\alpha^2\om_n} (\vA_n\sin\om_nt-\vB_n\cos\om_nt),
 % w_n=\sqrt{\frac{n^3}{\pi N^4}
\EEQ
and has two-point correlation function

\BEQ
\hspace{-2cm}
C^{AA}_{ab}(t-s)=\delta_{ab}C_{AA}(t-s),\quad
C_{AA}(t)=\frac{1}{4\pi N^2}\, \Re\,
\frac{1}{\sinh^2[(Z^2\alpha^2+it)/2N]}.
\EEQ

\subsection{Canonical momentum}

For large $\omega_n$, the coefficients of the $\vE$ field grow as $\om_n^{3/2}$, which
may cause numerical errors. To check whether this leads to numerical inconsistencies, we formulate  several presentations of the dynamics
where some of the integrations are performed analytically.

Firstly, the ``{\it canonical}\,'' dynamics reads

\BEQ
\dot \vr&=&\vp+\beta \vA+\beta^2\vf(\vr),
\quad \dot\vp=\vf(\vr) \, , \quad \nn\\
\dot \vq&=&\vp+\beta \vA+\beta^2\vf(\vr) \, , \quad {\vr=\vq}
\EEQ
When combined they reproduce (\ref{neqad}); notice that one does not need $\dot\vf$. 
The benefit is that at large $n$ (i.e. for high harmonics), $\vA$ has smaller coefficients than $\vE$,
inducing a better numerical stability.

The energy should not include the $\vA^2$ term, since it is already included in the renormalised mass.
With $V(r)=-\frac{1}{r}$ one has
\BEQ
\cE&=&\half \vp^2+\beta\vp\cdot\vA+V(r)
\nn\\ &=&\half\dot\vr^2+V(r)-\half\beta^2\vA^2,\qquad 
\EEQ
For the free particle one would have $\vp=\vp_0$ = constant, and $\langle\cE\rangle=\half\vp_0^2$.

\subsection{Grand canonical momentum}

We may proceed on this track.
Define $\vC=\int^t \d t\vA$, which amounts to

\BEQ
\vC(t)=\sum_{n=1}^\infty \sqrt{\frac{\Delta\om_n}{\pi \om_n}}  \left(-\vA_n\cos\om_nt+\vB_n\sin\om_nt\right) ,
\EEQ
and the canonical momentum   $\vp\equiv \int^t\d t\,\vf$ and  the canonical position $\vq=\int^t\d t\, (\vp+\beta^2\vf)$. Then, consider the dynamics
for $\vp$, $\vq$, using the physical position $\vr$,

%\end{document}

\BEQ \label{rCeq}
\dot \vp&=&\vf(\vr) ,\qquad  
\dot\vq=\vp+\beta^2\vf(\vr) , \nn\\
\vr&=&\vq+\beta\vC,
\qquad 
\dot\vr=\dot\vq+\beta\vA.
\EEQ
This is a ``{\it pure grand canonical}'' system of 6 first order equations, equivalent to the Newton problem (\ref{neqad}).
% The expression for $\vr$ is needed to evaluate $\vf$.
The initial conditions can be taken by neglecting $\beta$, so that $\vq(0)=\vr(0)$; $\vp(0)=\dot\vr(0)$.
The physical speed entering $\cE$ and $L$ is $\dot \vr=\dot\vq+\beta\dot\vC %+\beta^2\dot\vp
=\vp+\beta\vA+\beta^2\vf$; this extra evaluation of $\vA$ is needed at most once per orbit.

\subsubsection{Grand canonical momentum: second order differential equation}

In the above approach let us express $\ddot\vr=\vf-\beta\vE+\beta^2\dot\vf$ by a variable $\vs$ through 
the definition $\vq=\vs+\beta^2\vp$,

\BEQ \label{pqrs}
% \vq=\vs+\beta^2\vp,\qquad
% \EEQ \BEQ \label{rCeq}
%
{\dot \vp=\vf(\vr)}\, ,\qquad  
{\dot\vs=\vp}\, ,\qquad   
\vr=\vs+\beta^2\vp+\beta\vC,
%=\vs+\beta^2\dot \vs+\beta\vC}
\qquad 
\EEQ
They combine into a second order differential equation for $\vs$, 

\BEQ\label{seq}
\ddot\vs&=&\vf(\vr), \quad \vr=\vs+\beta\vC+\beta^2\dot\vs .
% \dot \vp(t)=\vf(\vr(t)),\qquad  \dot\vq(t)=\vp(t)+\beta^2\vf(\vr(t)),\qquad   \vr(t)=\vq(t)+\beta\vC(t)
\qquad
\EEQ

\subsubsection{Mixed grand canonical ensemble: Splitting up in low and high frequency components.}

If one splits into low frequencies $\om_n\le \om_\ast$ and high frequencies $\om_n>\om_\ast$
\BEQ 
\vC=\vC_l+\vC_h
\EEQ
one may define $\vu$ by
\BEQ
\vs=\vu-\beta\vC_l
\EEQ
and get the dynamics

\BEQ
\ddot\vu&=&\vf(\vr)-\beta\vE_l,\qquad \nn\\
 \vr&=&\vu+\beta^2\dot\vu +\beta\vC_h-\beta^3\vA_l\approx \vu+\beta^2\dot\vu +\beta\vC_h
\EEQ
The frequency components of $\vE_l$ for $\om_n\to0$ and $\vC_h$ for $\om_n\to\infty$ have small amplitudes.

\subsubsection{Fixed number of harmonics, moving number of frequency components}
\label{movingom-seq}

When working with a fixed number of harmonics, say $n_h=2.5$, and floating $\om_m=n_hk^3$, 
a change is needed in the equation of motion (\ref{seq}) when  the cutoff frequency $\om_m$ is updated because $k$ has changed noticibly.
Indeed, both the electron position $\vr$  and its speed $\dot\vr$ should not alter by the update.
Because of the form (\ref{pqrs}), it is natural to assume that  $\vs$, $\dot\vs$ and $\ddot\vs$ are continuous.

% \newpage

Let us start at time $t_0=0$ with $N_0$ terms in the sum. At a time $t_1$ this is changed to $N_1$, and successively to $N_{k+1}$ at times $t_{k+1}$ for $k=1,2\cdots$.
Let us write in the time interval
% $t_{k-1}<t<t_k$ and 
$t_k<t<t_{k+1}$
\BEQ
\vr&=&\vs+\beta^2\dot\vs+\beta\vC(t,N_{k})-\beta(\vu_{k}+\vv_{k}t),\quad \nn\\
\dot\vr&=&\dot\vs+\beta^2\ddot\vs+\beta\vA(t,N_{k})-\beta\vv_{k},
\quad (t_k<t<t_{k+1}) 
% \nn\\ \vr=\vs+\beta^2\dot\vs+\beta\vC(t,N_{k})-\beta\vw_{k}-\beta \vv_{k}(t-t_k),\quad 
% \dot\vr=\dot\vs+\beta^2\ddot\vs+\beta\vA(t,N_{k})-\beta\vv_{k}, \quad (t_k<t<t_{k+1})
\EEQ
At $t_k$ the fields $\vC$ and $\vA=\dot\vC$ make a step due to taking $N_{k}$ terms instead of $N_{k-1}$, viz. 

\BEQ
\Delta \vC_k&=&\vC(t_k,N_k)-\vC(t_k,N_{k-1}), \qquad \nn\\
\Delta \vA_k&=&\vA(t_k,N_k)-\vA(t_k,N_{k-1}), \qquad
\EEQ
Continuity of $\vr$ and $\dot\vr$ at time $t_k$ then requires

\BEQ 
\vu_k&=&\vu_{k-1}+\Delta \vC_k -\Delta \vA_kt_k,\qquad    
% \vw_k=\vw_{k-1}+\Delta \vC_k,\qquad    
\nn\\
\vv_k&=&\vv_{k-1}+\Delta\vA_k
\EEQ
One starts with $\vu_0=\vv_0=0$, so that

\BEQ
%\vw_k+(t-t_k)\vv_k=
\vu_k+\vv_kt=\sum_{l=1}^k\Delta\vC_l+\sum_{l=1}^k\Delta\vA_l(t-t_l), \qquad (t_k<t<t_{k+1}),
\EEQ
\BEQ
%\vw_k+(t-t_k)\vv_k=
\vu(t)+\vv(t)t=\sum_{l=1}^\infty\theta(t-t_l)[\,\Delta\vC_l+\Delta\vA_l(t-t_l)\,]
\EEQ

Since $\vr$ is continuous, so is $\ddot\vs=\vf(\vr)$, as assumed. 
% $\ddot\vr=\ddot\vs+\beta^2\dddot\vs-\beta\vE(t,N_{k})$
Also $\dddot\vs=\nabla\vf\ccdot\dot\vr$ will be continuous.
From $\ddot\vr=\ddot\vs+\beta^2\dddot\vs-\beta\vE(t,N_{k})$ it is seen that $\ddot\vr$ discontinuous,
as it is in the standard form of the Newton equation $\ddot\vr=\vf(\vr)-\beta\vE+\beta^2\nabla\vf\ccdot\dot\vr$.

The $\vv_kt$ shift in $\vr$ is possibly dangerous, since at large $t$ it may lead to large $|\vr|$.

\subsection{Mixed grand canonical ensemble}

In this scheme the high frequency components of the noise $\vC$ have decaying amplitude, but the small frequency part is strong, 
which does not do justice to the physics either.
To avoid this aspect, one may consider a combination of the two themes.
First, split up $\vA$ in ``smaller'' and ``greater'' frequency components,

\BEQ 
\vA&=&\vA_s+\dot \vC_g,\quad  \nn\\
\vA_s&=&\sum_{n=1}^{N_1}\sqrt{\frac{n}{\pi N^2}}  (\vA_n\sin\frac{nt}{N}+\vB_n\cos\frac{nt}{N}),
\quad \\ 
\dot \vC_g&=&\sum_{n=N_1+1}^\infty \sqrt{\frac{n}{\pi N^2}}  (\vA_n\sin\frac{nt}{N}+\vB_n\cos\frac{nt}{N}), \\
 \EEQ

% \newpage

\noindent
Consider the ``{\it mixed grand canonical}'' dynamics %on $\vC$ and $<$ on $\vA$)
for the canonical momentum $\vp$, a modified canonical position $\vq$ and the physical position $\vr$,

\BEQ\label{CAP1}
{\dot \vp=\vf(\vr),}\qquad 
{\dot\vq=\vp+\beta \vA_s+\beta^2\vf(\vr),}\qquad 
{ \vr=\vq+\beta\vC_g} %+(1-c)\beta^2\vp}
\EEQ
 This implies the physical momentum $\dot \vr=\vp+\beta(\vA_s+\dot\vC_g)+\beta^2\vf$.

\noindent
These also combine to Eq. (\ref{neqad}). The benefit is that both $\vA_s$ and $\vC_g$ are well-behaved sums with maximal coefficients at $n=N_1$
for $\vA_s$ and at $n=N_1+1$ for $\vC_g$.
The most logical choice is the fixed case $N_1$. One may choose $N_1=N$; even better is $N_1=(2/3)^{3/2}N=0.5443N$, which 
puts $\omega_{N_1}=k_m^3$ at  $\cE_m=-\half k_m^2=-\frac{1}{3}$ where $P(\cE)$ is maximal.
It would not change much to take just $N_1=\half N$.

\subsubsection{Final dynamics: changing $N_1$ and $N_2$}

The present argument remains valid for numerical approaches, where we have to approximate $\vC\equiv \vC_g$ as a finite sum, $\vC_g=\sum_{N_1+1}^{N_2}\vC_n$, 
and $N_1$ and $N_2$ are updated simultaneously.
Let us assume that this covers $n_h+\half$ harmonics of the orbit, with $n_h=2$ or $4$, or $\cdots$.
At the initial time we set $N_2=(n_h+\half )k^3N$, next to $N_1=k^3N$.

At some later time $t'$ where $k$  has evolved to some $k'$ we may wish to update not only $N_1$ but also $N_2$, to become 
$N_1'=k'{}^3N$ and $N_2'=(n_h+\half )k'{}^3N$.
This change is also covered in the above formulae, where now $\vC$ involves limits $N_1+1$ and $N_2$ before $t'$ while
the update $\vC'$ involves limits $N_1'+1$ and $N_2'$ after $t'$.
Likewise, $\vA$ involves limits $1$ and $N_1$, and  
 $\vA'$ involves limits $1$ and $N_1'$.

All by all,  the dynamics can general be described by

\BEQ \label{CAP2}
\dot \vp&=&\vf(\vr)\, ,
%\quad 
\,\, \dot\vq(t)=\vp+\beta^2\vf +\beta [\vA'(t)+\delta\vA] \, , 
\qquad 
\nn \\
 \vr(t)&=&\vq(t)+\beta[\vC'(t)+\delta\vC]
\EEQ

In the initial period, one just has $\delta\vA=\delta\vC=0$
  while $\vA'=\vA$, $\vC'=\vC$ are given by

\BEQ
\vA &=&\sum_{n=1}^{N_1 }\sqrt{\frac{n}{\pi N^2}}  (\vA_n\sin\frac{nt}{N}+\vB_n\cos\frac{nt}{N}),
\qquad \nn\\
\vC &=&\sum_{n=N_1 +1}^{N_2 } \sqrt{\frac{1}{\pi n}}   (-\vA_n\cos\frac{nt}{N}+\vB_n\sin\frac{nt}{N})
\EEQ
  After the first change of $N_1$ and $N_2$ one 
works with the updates $\vA'$ and $\vC'$, 
which involve $N_1'$ and $N_2'$, rather than $N_1$  and $N_2$, respectively.
%given by \BEQ
% \vA' &=&\sum_{n=1}^{N_1'}\sqrt{\frac{n}{\pi N^2}}  (\vA_n\sin\frac{nt}{N}+\vB_n\cos\frac{nt}{N}), \qquad  \nn\\
% \vC'&=&\sum_{n=N_1'+1}^{N_2'} \sqrt{\frac{1}{\pi n}}   (-\vA_n\cos\frac{nt}{N}+\vB_n\sin\frac{nt}{N}) \EEQ
Matching at $t'$ yields

\BEQ \label{CAP2}
%{}
\delta\vA&=&\vA(t')-\vA'(t')+\dot\vC(t')-\dot\vC'(t') ,\qquad \nn\\
%{}
\delta\vC&=&\vC(t')-\vC'(t')
\EEQ
For subsequent changes of $N_1$, $N_2$ one repeats this schedule. One must add the new shifts  to the previous ones, 

\BEQ \label{CAP2}
\delta\vA_{\it new}&=&\delta\vA_{\it old}+\vA(t')-\vA'(t')+\dot\vC(t')-\dot\vC'(t') \, ,\qquad
\nn\\
\delta\vC_{\it new}&=&\delta \vC_{\it old}+\vC(t')-\vC'(t')
\EEQ
which amounts in total to

\BEQ \label{CAP2}
\delta\vA&=&\sum_{t'<t}[\vA(t')-\vA'(t')+\dot\vC(t')-\dot\vC'(t')] \nn\\
\delta\vC&=&\sum_{t'<t}[\vC(t')-\vC'(t')] .
\EEQ

These forms have been applied to test the results of our simulations.

\section{Conjecture for the ground state phase space density} 

For a dynamics with weak noise the stationary distribution in phase space must be a function of the conserved quantities, 
here the seven parameters $\cE$, $\vL$ and $\varepsilon $. 
They contain the scalars $\cE$, $L$ and $\eps$,  while the coordinate-invariant inner product ${\it \vL} \cdot \bvarepsilon$ vanishes. 
Because of the relation (\ref{epsEL}), two of the scalars are independent.

A conjecture for the phase space density of several states of the relativistic H-atom has been made by one of us \cite{NRelHatom06}.
Here we restrict ourselves to the ground state in the non-relativistic limit. 
The conjecture reduces to

\BEQ\label{Pmu=}
\hspace{-2.3cm}
P_{\vp\vr}(\vr,\vp)=f(\cE(\vr,\vp),L(\vr,\vp)); \quad 
f(\cE,L)
= \frac{2Le^{2/\cE}}{\pi^3 |\cE|^3}
=\frac{2}{\pi^3}\,LR^3e^{-2R}
,\quad R=-\frac{1}{\cE}.
\EEQ
The first task is to verify that the ground state density emerges after integrating over momenta.
At given $\vr$ one can take the $p_z$-axis along ${\bf r}$, so that 

\BEQ
\vp =p(\sin\mu\cos\nu,\sin\mu\sin\nu,\cos\mu),\quad p=\sqrt{\frac{2(R-r)}{rR}},
\EEQ
with  $r\le R\le \infty$,  $ 0\le\mu\le\pi$, $ 0\le\nu\le 2\pi$. 
The  volume element reads

\BEQ
\d^3p=\d p\d\mu\d\nu\,p^2\sin\mu=
 \d R\d \mu\d \nu\,\,\sqrt{\frac{2(R-r)}{rR^5}}\,\sin\mu.
\EEQ
Since $L=pr\sin\mu$,  Eq.  (\ref{Pmu=}) indeed reproduces the QM result, viz.

\BEA
\hspace{-.5cm}
P_\vr(\vr)&=&\int\d ^3p\, P_{\vp\vr}({\bf r},{\bf p})=\frac{4}{\pi}\int_r^\infty\d R(R-r)e^{-2R}
 =\frac{e^{-2r}}{\pi}. \EEA
 This can indeed be written as

\BEA
P_\vr(\vr)&=&\psi_0^2(r)Y_{00}^2(\theta,\phi),\quad \psi_0(r)=2e^{-r},\quad Y_{00}(\theta,\phi)=\frac{1}{\sqrt{4\pi}} ,
\EEA
and leads to $P_r(r)=r^2\psi_0^2(r)=4r^2 e^{-2r}$ with  normalisation $\int_0^\infty\d r\, P_r(r)=1$.

For $P_{\cE L}(\cE ,L)$ we have the definition

\BEQ
P_{\cE L}(\cE ,L)
&=&\int\d^3r\int\d^3p\,\delta(\underline \cE -\cE )\delta(\underline L-L)P_{\vp\vr}(\vp,\vr)
\nn\\
% &=&R^2P_{\vp\vr}(\cE ,L)\int\d^3r\int\d^3p\,\delta(\underline R-R)\delta(\underline L-L) \\ %P_{\cE L}(\cE ,L)
&=&4\pi f(\cE ,L)\int\d r\,r^2\,\int_0^{2\pi}\d\nu\int_0^{\pi}d\mu \sqrt{\frac{2(R-r)}{rR}}\,\nn
\\ &&\times \sin\mu\,\delta\left(r\sqrt{\frac{2(R-r)}{rR}}\sin\mu-L\right) 
\EEQ
Hence, %integrating over $\nu$ and 
taking into account the contributions from $\mu=\bar\mu<\half\pi$ and from $\mu=\pi-\bar\mu$,

\BEQ
P_{\cE L}(\cE ,L)
% &=&P_{\vp\vr}(\cE ,L)4\pi\int\d r\,r^2\,4\pi\int_0^{\half\pi}d\mu \sqrt{\frac{2(R-r)}{rR}}\,\sin\mu\,\delta\left(r\sqrt{\frac{2(R-r)}{rR}}\sin\mu-L\right) \nn\\ &=&16\pi ^2 P_{\vp\vr}(\cE ,L)\int\d r\,\frac{r\,\sin\bar\mu}{\cos\bar\mu} \nn\\
&=&16\pi ^2 f(\cE ,L)\int_{r_-}^{r_+}\d r\,\frac{rL\sqrt{R/2}}{\sqrt{rR-r^2-\half L^2R}}\nn
% &=&16\pi ^2 P_{\vp\vr}(\cE ,L)\int_{r_-}^{r_+}\d r\,\frac{rL\sqrt{R/2}}{\sqrt{\frac{1}{4}R^2-\half L^2R-(r-\half R)^2}}\nn\\
\EEQ
Expressing $\kappa=kL$, that lies between 0 and 1, as 
\BEQ
\kappa =\frac{L}{L_{\rm max}}=\frac{L}{\sqrt{R/2}}=kL=\sqrt{1-\eps^2},\qquad
% r_\pm=\half R(1\pm\eps),
\EEQ
and using that $ r_\pm=\half R(1\pm\eps)$,
this reduces to

\BEQ
P_{\cE L}(\cE ,L)
%=P_{\vp\vr}(\cE ,L)4\sqrt{2}\pi ^3LR^{3/2}
=8\sqrt{2}\frac{L^2}{|\cE |^{9/2}}e^{-2/|\cE |}\,,
%=8\sqrt{2}L^2R^{9/2}e^{-2R}\,,
%\qquad 0<L<L_{\rm max},
\EEQ
where $L\le L_{\rm max}$. Because the latter depends on $\cE $, the result does not factorize.
However, since both $\eps$ and $\kappa$ lie between 0 and 1,
 the weight $P_{\cE L}(\cE ,L)\d \cE \d L$ can be factored in the forms
$P_\cE (\cE )\d \cE \,P_\eps(\eps)\d\eps$ and $P_\cE (\cE )\d \cE \,P_\kappa(\kappa)\d\kappa$, where

%\BEQ\label{PRkappa=} P_{R\kappa}(R,\kappa)\d\kappa\d R=P_{\cE L}(\cE ,L)\d \cE \d L
% =\d(\kappa^3)p_\cE (\cE )\d \cE =\d(\kappa^3)p_R(R)\d R=\d(\kappa^3)\frac{4}{3} R^4e^{-2R}\d R,\EEQ

\BEQ\label{PRkappa=}
P_{\cE }(\cE )&=&\frac{4}{3|\cE |^6} e^{-2/|\cE |}\,,\qquad (-\infty<\cE <0), \nn\\
%{P_{R}(R)\d R=\frac{4}{3} R^4e^{-2R}\d R}\, ,\qquad 
P_\eps(\eps)&=&3\eps\sqrt{1-\eps^2} ,\qquad (0\le\eps<1), \\
P_\kappa(\kappa)&=& 3 \kappa^2 ,\qquad \quad \qquad (0<\kappa\le 1).\nn
\EEQ

For numerical simulation of an ensemble of orbits,
a properly distributed set of initial values can be gotten as follows.
Choose randomly two independent random numbers $u_1$ and $u_2$
%,  uniformly distributed 
between 0 and 1
and equate $R$ and $\kappa$ from %, $\eps$ and $L$ from 

\BEQ 
 (1 + 2 R + 2 R^2 + \frac{4}{3} R^3 + \frac{2}{3} R^4)e^{-2R}=u_1,\qquad \kappa=u_2^{1/3},\quad 
 \EEQ
 and from them the other parameters that characterise the orbit.
 
% \BEQ
 %\eps=\sqrt{1-\kappa^2},\quad L=\kappa\sqrt{\frac{R}{2}}=\frac{\kappa}{k},\quad E=-\frac{1}{R},\quad 
% k=\sqrt{\frac{2}{R}},\quad r_\pm=\half R(1\pm\eps)\EEQ
A uniform distribution of $u_1$ and $u_2$ values produces the desired probability density, viz. $\d u_1\d u_2=-\d\kappa\d R \,P_{\kappa R}(\kappa, R)$.
For an ensemble of initial conditions, the task is to see whether this ensemble is dynamically stable.
% From $\kappa$ and $R$ we have $L=\kappa\sqrt{R/2}$ and $E=-1/R$.%  and then $\eps^2=1-\kappa^2$. 
The initial orbit has perihelion and aphelion $r_\mp$ and can start at either of them.
% \BEQ -\d u_1\d u_2 =(3\lambda^2\d\lambda)(\d R\,\frac{4}{3} R^4e^{-2R}) =(3\lambda^2\d\lambda)(\d E\,\frac{4}{3E^6} e^{2/E})\EEQ
% produces the equilibrium density.

The distribution of the physical momentum is 

\BEQ 
P_p(p)=\int\d ^3r\,P_{\vp\vr}(\vp,\vr)
\EEQ
Taking and spherical coordinates with the $z$-axis parallel to $\vp$, we have $L=pr\sin\theta$ and then from $\half p^2-1/r=-1/R$

\BEQ
P_p(p)
=\frac{2p}{\pi}\int_0^{2/p^2}\d r\, r^3R^3e^{-2R}
=\frac{2p}{\pi}\int_0^\infty \d R\,\frac{R^6}{(1+\half p^2R)^5}e^{-2R}
\EEQ
which is also properly normalised. Indeed, its integral over the $3d$ momentum can be written as

\BEQ
%4\pi\int_0^\infty\d p\,p^2P(p)=
16\int_0^\infty\frac{\d x\,x}{(1+x)^5}\int_0^\infty\d R\, R^4e^{-2R}%=16B(2,3)\frac{4!}{2^5}
=1.
\EEQ
The limiting behaviours of $P_p(p)$ are $45p/4\pi$ for $p\to 0$ and $16/\pi p^9$ for $p\to\infty$.

The Wigner function is
generally defined as

\BEQ
W(\vp,\vr)=\frac{1}{(2\pi)^3}\int\d^3s\,\psi_0(\vr-\half\vs)\psi_0(\vr+\half\vs)e^{i\vs\cdot\vp}
\EEQ
It implies $W_p(p)=\int\d^3\vr\,W(\vp,\vr)$, which yields $W_p(0)=1$, to be compared with $P_p(0)=0$. The distribution functions are physically different:
the  Wigner function deals with the statistical momentum and $P_{\vp\vr}$ with the
instantaneous momentum. Except for Gaussian distributions, $W$ will have negative parts, while $P_{\vp\vr}$ is always nonnegative.

\newcommand{\Nharmonics}{$N_{\rm harm}$}

\section{Implementing the algorithm in OpenCL}

We did make extensive use of GPGPU computing by writing our code in C++/OpenCL. This led to a factor $10^2$ improvement in processing speed with respect to 
a normal auto-vectorized single core C++ implementation, which allowed us to simulate on the order of $10^7$ modes in real time, allowing us to tackle the 
problem in 3D without making use of window approximations as was done in previous simulations (Cole\&Zou 2003). In the next sections we will elaborate on 
how this led to vastly different results with respect to previous modeling work (Cole\&Zou 2003). 

We ran the simulations on a state of the art PC, consisting out of an Intel Core i7 2600k overclocked to 4.6 GHz (Core i7 4770k+ equivalent in performance), 
together with 16 GB of RAM.  In our earlier simulations we used an AMD HD6970 GPU, which was later upgraded to an AMD R9-290X GPU. 
This GPU delivers 5.6 TFLOPS of single precision floating point performance and 350 Gigabytes per second of memory bandwidth.

Solving the equation of motion is done with the Runge-Kutta fourth order ODE integration scheme. The most demanding part of this constitutes the summation 
of all modes of the E-field. This is where we use OpenCL. Since all these modes are independent of each other, we can sum them in parallel on the GPU. 
Basically we reduce this tremendous ($10^7$) sum into $10^4$ sub sums, ordered in groups of size 256 sharing local memory on the GPU,  divided over 
1536/2816 stream processors. The final reduction step of the $10^4$ sub sums is done on the CPU. This costs no extra time and has the advantage that the 
final reduction step can be done with double precision, while on the GPU it is done with single precision. We confirmed that our required numerical precision 
was met by comparing the parallel OpenCL GPU reduction with a normal double precision C++ CPU only reduction.

For the Runge-Kutta fourth order algorithm to remain stable on our timescales we found out that we need approximately $600-2000$ iterations per orbit for high eccentricities. 
In our code we used at least 4000 iterations per orbit, since solving the ODE, if the E-field is known, is computationally inexpensive. The catch is that we 
can not update this E-field so often, because every update involves a $10^7$ sized reduction. For this reason we update the field only 10 times per period 
for the highest frequency mode in the spectrum. In between we use a 4$^{th}$ order Lagrange Polynomial for interpolation to calculate the E-field.

Since the electron's energy can drop below the memory limit of our simulation ($\cE=-1.6$), in such a case we artificially increase its energy by giving it a `push' parallel to the electron's velocity vector. Naively, this shouldn't constitute a problem, since according to the conjecture of previous section the electron should stay out of this regime 99\%+ of time. We tested that the electron can drop to very low energies like $\cE=-4.0$, but the electron always seems to recover from this regime. However, we do observe that in this regime the angular momentum and eccentricity can change fast, possibly biasing our final results.

We ran our simulations for around $1-5\,10^6$ orbits for the different sets of harmonics (with or without a smooth window). 
This is many times more than the total run time of Cole\&Zou 2003. See table 1 for the precise run times.

\subsection{Results}

\subsubsection{Moving cutoff.}

Our first simulations utilised a `moving' cutoff for the electric field.  We took the cutoff at \Nharmonics \ times the orbital frequency. We updated this cutoff frequency in increments of 20\% as our orbit changed. This has the advantage that we do not introduce discontinuities in the Abraham-Lorentz equation of motion (10). For Z=3 experience shows that energy of the electron varies by a few percent every orbit and thus the electric field is updated every $10^2$ orbits.

Our most promising results were multiple simulations for \Nharmonics \ = 2.5 and $N=1.5*10^6$ using either $Z=1$ or $Z=3$ (see figures 1-6). 
Initially it seemed that we obtained a stable solution, but  instabilities, which eventually led to ionization of the atom, 
developed on timescales of the order $10^7 t_0$ \ for \Nharmonics \ = 2.5 and $Z=3$. Higher harmonics (4.5 and 6.5) are unstable on even shorter timescales ($10^6-10^5 t_0$). 
We define ionisation as the moment when the electron stays above $\cE=-0.05$ for a duration for at least $10^7 t_0$. The moment of ionisation is cut out
of the subsequent plots.

\begin{figure}[h!]
\label{fig1}
% \centerline{ \includegraphics[width=8cm]{SN-1987A-dust-mass.png}}
\centerline{ \includegraphics[width=8cm]{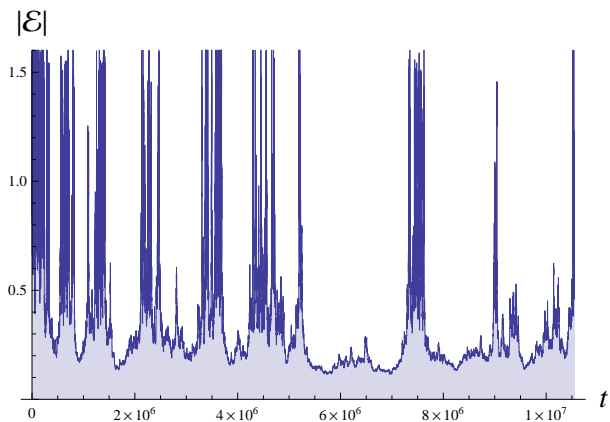}}
\caption{Energy of the electron as function of time for $Z=3$, in Bohr units.}
\end{figure}

\begin{figure}[h!]
\label{fig2}
% \centerline{ \includegraphics[width=8cm]{SN-1987A-dust-mass.png}}
\centerline{ \includegraphics[width=8cm]{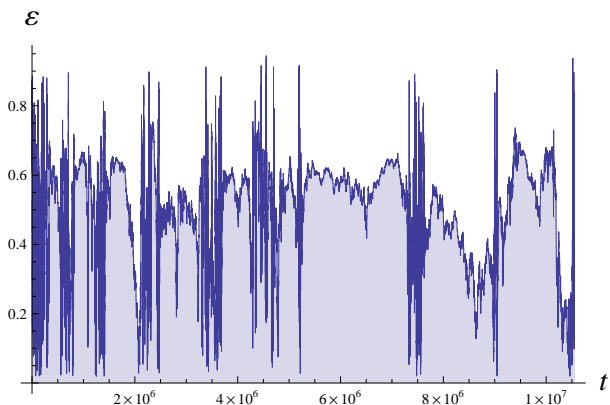}}
\caption{Eccentricity of the electron orbit as function of time for $Z=3$, in Bohr units.}
\end{figure}

\begin{figure}[h!]
\label{fig3}
% \centerline{ \includegraphics[width=8cm]{SN-1987A-dust-mass.png}}
\centerline{ \includegraphics[width=8cm]{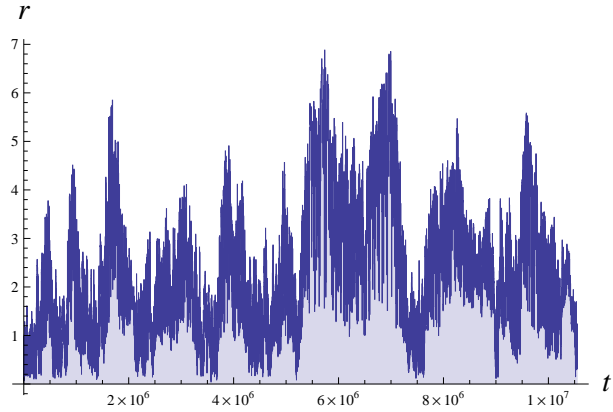}}
\caption{Radius of the electron orbit as function of time for $Z=3$, in Bohr units.}
\end{figure}

\begin{figure}[h!]
\label{fig4}
% \centerline{ \includegraphics[width=8cm]{SN-1987A-dust-mass.png}}
\centerline{ \includegraphics[width=8cm]{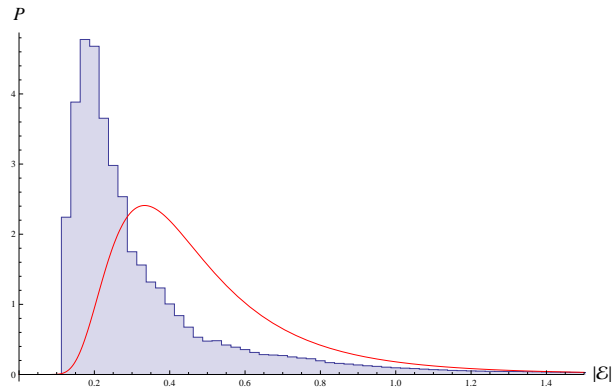}}
\caption{Normalised histogram of the electron energy for $Z=3$ versus the conjecture, in Bohr units.}
\end{figure}

\begin{figure}[h!]
\label{fig5}
% \centerline{ \includegraphics[width=8cm]{SN-1987A-dust-mass.png}}
\centerline{ \includegraphics[width=8cm]{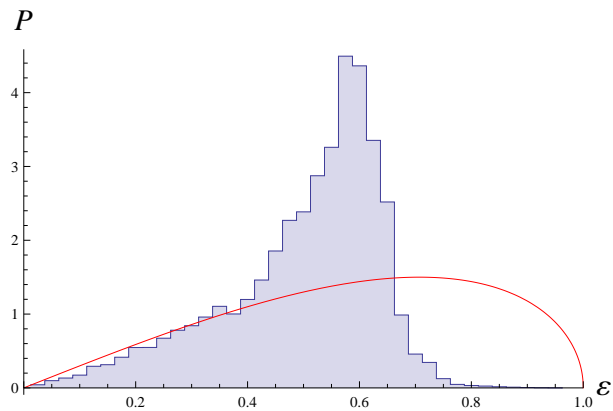}}
\caption{Normalised histogram of the eccentricity of the electron orbit for $Z=3$ versus the conjecture, in Bohr units.}
\end{figure}

\begin{figure}[h!]
\label{fig6}
% \centerline{ \includegraphics[width=8cm]{SN-1987A-dust-mass.png}}
\centerline{ \includegraphics[width=8cm]{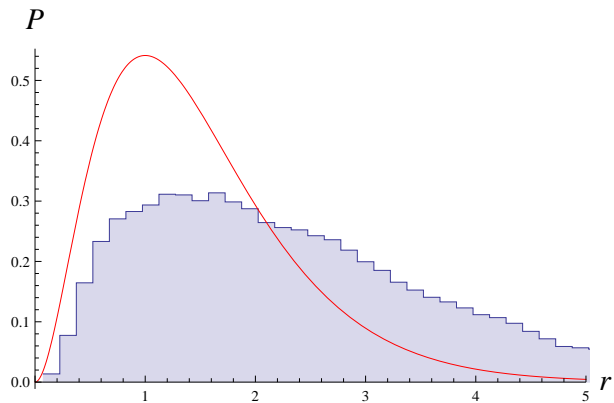}}
\caption{Normalised histogram of the radius of the electron orbit for $Z=3$ versus the conjecture, in Bohr units}
\end{figure}

\begin{center}
\noindent     
\begin{tabular}{||r|r|r||}  
 \hline \hline    
property & value & duration (s) \\ 
 \hline\hline
$ t_{total}$ & $1.2\,10^7\,t_0$ & $3.2\,10^{-11}$ s \\ \hline
$t_{damp}$&  $4.3\,10^5\,t_0$ & $1.2\,10^{-12}$ s \\ \hline
$N_{orbit}$ &  $1.9 \, 10^6$ &  $1.0 \ 10^6$  \\  \hline
$N_{damp}$ & 28 & \\ \hline \hline
\end{tabular}     
\end{center}     
{\small  Table 1: Duration and number of orbits for our simulation with  2.5 harmonics and $Z=3$. The classical period of a characteristic orbit is 
the Bohr period $t_0=2\pi \tau_0$. The number of orbits is first given as the total duration divided by $t_0$; 
its second entry is the actual number of orbits in the simulation. $N_{damp}$ is the duration of the simulation expressed in damping times.}

In these time series (figures 1,2,3) we see a rapid fluctuation of the energy and radius, while the eccentricity and angular momentum remain stable on 
longer timescales.  Contrary to our expectations and the ones of Cole\&Zou 2003, we fail to reproduce the 1s wavefunction on such long timescales. 
Comparing the angular momentum, eccentricity and energy distributions to our conjecture we fail to see high eccentricities, while we confirmed that
our simulation can  accurately handle eccentricities of at least $\sim0.99$. The cause of the lack of these higher eccentricities remains unanswered so far. 

 One of the striking results from Cole\&Zou 2003 was that they observed perfectly circular orbits. This was not true in our case and in strong contradiction 
 with the previously discussed conjecture. Our explanation for this discrepancy is that Cole\&Zou 2003 did run the simulation too short and thus missed 
 variation of the angular momentum and eccentricity on longer timescales. Cole\&Zou 2003 furthermore summed 11 simulations with the same initial 
 conditions (circular orbit) but different seeds, which clearly is not a valid approach anymore and leads to wrong results (i.e., they chose a timescale much 
 shorter than the timescale on which the eccentricity changes significantly). Furthermore a window approximation of 5\% around the first harmonic of the 
 E-field was used. We have verified that this leads to wrong results for even a single harmonic, since the E-field components outside the `resonant' 
 window of several percent seem to influence the angular momentum and eccentricity distributions on longer timescales.
If we follow Cole\&Zou by taking the data for $r(t)$ of Fig. 3 up to the smaller time $2.5 \ 10^5/2\pi=16,000$ orbits with a nearly 
circular orbit as initial condition, we get as radius distribution the data presented in Table 2:
  
  \begin{figure}[b!]
\label{figSN87}
% \centerline{ \includegraphics[width=8cm]{SN-1987A-dust-mass.png}}
\centerline{ \includegraphics[width=8cm]{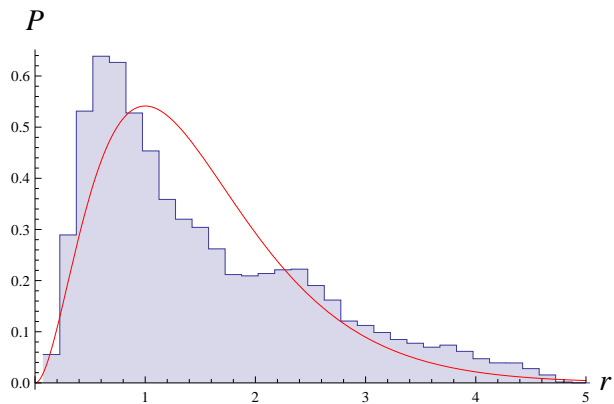}}
\caption{Normalised histogram of the radius of the electron orbit for $Z=3$ versus the conjecture, in Bohr units. 
The data are taken from Fig. 3 for times up to $10^5t_0$, corresponding to an estimated number 16,000 of orbits, 
about 100 times shorter than in Fig. 6. }
\label{figSN87}
\end{figure}

% $$ Cole's times (Z=1):  Total: 11*1.5*10^-11 s N_orbit: 11*1.0*10^5 T_damp: 9.36*10^-11 s N_damp: 11*0.17 $$

\begin{center}
\noindent     
\begin{tabular}{||r|r|r||} 
 \hline \hline    
property & value & total \\ 
 \hline\hline
$ t_{total}$ & $ 11\times 6.21\,10^5t_0$  & $11\ast 1.5\, 10^{-11}$ s \\ \hline
$t_{damp}$& $3.8\,10^{6}t_0$ & $9.36\,10^{-11}$ s \\ \hline
$N_{orbit}$ & $11\times 1.0\, 10^5$ &   \\ \hline
$N_{damp}$ &$11\times 0.16 $&  \\ \hline \hline
\end{tabular}     
%v\caption{aabb}
\end{center}     
% \centerline
{Table 2: Duration and number of orbits for the Cole\&Zou simulation with $Z=1$ and a 5\% window around the first harmonic in 2D. }
%\centerline
{The factor 11 is the number of different runs.}

\subsubsection{Fixed cutoff}

After we upgraded to a more powerful GPU, we ran simulations with a fixed cutoff on the frequency spectrum of the the random electric fields, 
so that we could keep the field and time step the same during the whole 
simulation. This cutoff was set at \Nharmonics \ = 1.5 harmonics for an energy of $\cE=-1.6$ so that \Nharmonics \ = 52 harmonics occur at energy of $\cE=-0.15$. 
Multiple configurations led to ionization in a rather short amount of time (10.000 orbits) . We observe that the energy of the electron
goes to zero,  while its eccentricity increases, before it ionizes (see figures 7-9) .

\begin{figure}[h!]
\label{figSN87}
% \centerline{ \includegraphics[width=8cm]{SN-1987A-dust-mass.png}}
\centerline{ \includegraphics[width=8cm]{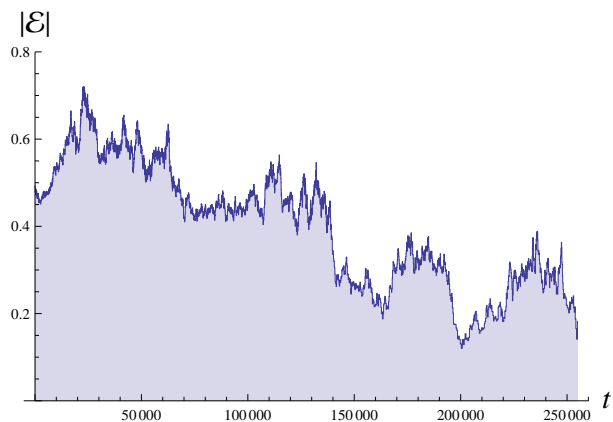}}
\caption{Energy as function of time for $Z=1$ with a fixed cutoff exposing the trend towards ionisation at $\cE=0$, $\eps=1$. 
The time window is 45 times shorter than in Figures 1, 2 and 3.}
\label{figSN87}
\end{figure}

\begin{figure}[h!]
\label{figSN87}
% \centerline{ \includegraphics[width=8cm]{SN-1987A-dust-mass.png}}
\centerline{ \includegraphics[width=8cm]{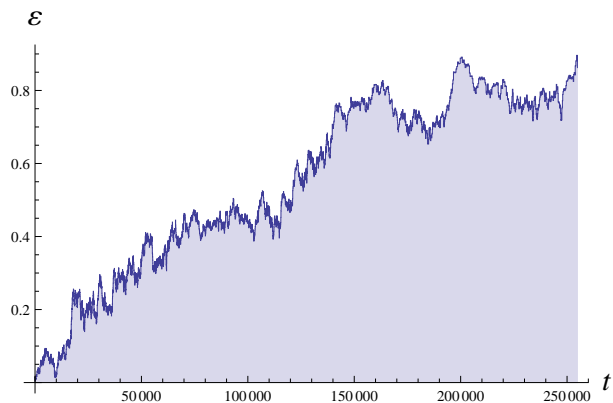}}
\caption{Eccentricity as function of time for $Z=1$ with a fixed cutoff exposing the trend towards ionisation at $\cE=0$, $\eps=1$. 
The time window is 45 times shorter than in Figures 1, 2 and 3.}
\label{figSN87}
\end{figure}

This is unphysical and raises questions about our numerical precision in the fixed cutoff case. Since we use  double-precision, 
the machine precision is not our limiting factor.  When the electron is close to the lower energy limit from the conjecture ($\cE=-0.15$), though, we have 
to include 52 harmonics, in order that at the lower energy threshold of $\cE=-1.6$ we retain the promised 1.5 harmonics. 
This means that integrated strength of the 52$^{th}$ harmonic is  $\sim 2700$ stronger 
 than the integrated strength of the first harmonic. Hence we must keep our 4$^{th}$ order interpolation error very small so that we correctly 
 represent the integrated strength up to the first harmonic, which should be one of the most dominant ones, with the higher harmonics statistically averaging out. 
 Our tests turned out that our `numerical' error in this case can reach up to 20\% of the integrated strength of the first harmonic. 
 Since this is a small statistical error, we expect it to average out and not lead to too much systematic disruption of our data. We tested this further by experimenting with an exponential
cutoff, which ranged in values from 1 at the lowest frequency in our spectrum to 0.1 at the cutoff frequency. No improvement of the electron stability was observed.

\section{Summary and outlook}
 
 In this work we have considered the hydrogen ground state in Stochastic Electrodynamics (SED). The approach was made tractable 
 by replacing the Gaussian random field, sums over $3D$ momenta, by $1D$  sums over frequencies 
 with amplitudes chosen such that they reproduce the same correlation function in the limit where the frequency mesh vanishes ($N\to\infty$).
 
Using vastly improved computational resources compared to a decade ago and the possibility to parallelise the code into OpenCL,
we could simulate up to much longer timescales than in Cole\&Zou 2003.
We considered simulations both with a fixed cutoff and with a moving cutoff frequency. The results show that even for a simple problem consisting out of a moving cutoff
at 2.5 times the electron angular frequency, the promising results from Cole\&Zou 2003 are not valid anymore on longer timescales. 
This result is robust against several variations that we implemented, such as the `grand canonical',  `mixed grand canonical' and other schemes of the dynamics 
presented in section 2, which allowed to test our numerical reproducibility. If we include more harmonics or if
we use a fixed cutoff, the solution is unstable on even shorter timescales. Furthermore, the solutions fail to reproduce the 1s ground state correctly even prior to
ionisation.

% \section{Conclusion}

We suspect that the H atom suffers from the same problems as the free particle \cite{PenaFreeParticle}.
The energy $e^2\vA^2/2mc^2$ is in quantum mechanics perceived as a renormalisation $\delta m_ec^2$ of the electron rest energy.
In SED it is a dynamical energy, transferred  by the $\vp\cdot\vA$ term in the Hamiltonian,
 that gives energy to the electron when time progresses. Its energy scale $\alpha m_ec^2$
is much larger than the Rydberg energy $\half \alpha^2 m_ec^2$.
It seems to us that this transfer of physical energy from the field to kinetic plus potential energy in the H atom causes the ionisation.
One may wonder whether a compensation mechanism for this transfer exists.

An improvement to our approach is to include the magnetic  field, the weak spatial dependence of the electric field
and the effect of the spin-orbit coupling. They will be considered in the near future. 
Relativistic effects are important for the problem
provided the electron comes close to the nucleus  \cite{Boyer2003,Boyer2004}.
Otherwise they appear in the structure of SED as small mechanical corrections to the Kepler problem, 
which makes us suspect that they do not significantly alter the present findings.

\section*{Appendix: Parallelising the code into OpenCL}
OpenCL is an extension to C++ that makes it possible to parallelise the summation in equation (\ref{EadtB}). 
Normally the summation of the modes is performed within for loops, where all elements are summed serially on a single CPU.
GPUs though are much faster ($\sim30\times)$ and share a much higher memory bandwidth ($\sim20\times$) than CPUs. They are built up out of thousands
 of stream processors. Each of these stream processors is much weaker than a single CPU core (usually 2 to 4 cores per CPU), but taken together they are tens to 
 hundreds of times faster than a CPU. To utilise this strength we used OpenCL to program our GPU. 
In the OpenCL paradigm our GPU is called a compute device. Since we possess a single GPU, we utilise only one compute device. This compute device posseses 
44 compute units, which are subdivided over 4 16-wide SIMDs (Single instruction, multiple data). 
Thus there are $64\times4\times16=2816$ processing elements. Each of these compute units can process up to 40 wavefronts of GPU specific size 
64 simultaneously, only limited by the register (GPR) and local memory size. Processing multiple wavefronts of data is done to hide memory latency.

The sum in equation (15) is then summed by all of these processing elements using parallel reduction. The global work size is defined by the total number of 
elements to sum. These elements are summed in workgroups of size 256, i.e., 4 times the wavefront size for an AMD GCN (Graphics core next) GPU. 
These workgroups share local memory, such that every work item reads in its value from the global memory and copies it to the local memory, where it 
is summed in 8 steps ($2^8=256$) within a workgroup. During the first step the first 128  work items are summed with the last 128 work items in pairs of two. 
In this way the sum is reduced by a factor of 2 each step.
This reduces the total sum by a factor of 256, after which the remainder is copied to pinned RAM memory via a PCI Express bus and summed by the CPU. 
Overall, the performance improvement amounts to a factor of 50-300 depending on the exact combination of CPU and GPU.

The OpenCL code is available upon request.

%Table 2.5 harmonics Our times (Z=3):
% Total: $1.57\,10^7t_0$ OR $4.23\,10^{-11}$ s $N_{orbit}$:  $2.49\,10^6$ $T_{damp}$:  $4.28\,10^5t_0$ OR $7.26\,10^{-12}$ s $N_{damp}$: 37

\section{Acknowledgments}

It is a pleasure to thank Erik van Heusden for much discussion.

\vspace{5mm}

\newcommand{\asas}{Astron. \& Astrophys}
\newcommand{\apj}{Astroph. J.}
\newcommand{\pasp}{pasp}

\end{document}